# Principles and dynamics of quantum mechanics

Spyros Efthimiades
*Department of Natural Sciences, Fordham University, New York, NY 10023*
*email*: sefthimiades@fordham.edu

The fundamental principle of quantum mechanics is that the probabilities of physical outcomes are obtained from the intermediate states and processes of the interacting particles, considered as happening concurrently. When the interaction is described by a potential, the total energy of the particle is equal to its total kinetic plus potential energies. We derive the Schrodinger and Dirac equations as the conditions the wavefunction must satisfy at each point in order to fulfill the corresponding energy equation. In our approach quantum theory is tangible, experimentally justified and theoretically consistent.

## I. INTRODUCTION

There is a dichotomy in the traditional approach to quantum dynamics. We obtain the probabilities of outcomes of simple boundary interactions, i.e. double slit experiment, from the interference of the intermediate particle waves. However, when a particle interacts with a potential, the particle wavefunction is calculated from the postulated Schrodinger equation that is considered to play "a role analogous to Newton's second law" and which "cannot be derived from simpler principles but it is the fundamental principle of quantum mechanics".[1-2] This axiomatic stance causes quantum dynamics to have an abstract basis and disjointed parts. In addition, the physical origin of the wavefunction is unclear since it is the solution of a postulated equation.

In our approach, the probabilities of physical outcomes are obtained, in all cases, from the superposition of the intermediate particle waves and interaction processes, considered as happening concurrently. When a particle interacts with a potential, the total energy of the particle is equal to its total kinetic plus potential energies. We then derive the Schrodinger equation as the unique condition the wavefunction must satisfy at each point in order to fulfill the energy equation.

First, we will discuss particle waves and wavefunctions. Next, we derive the Schrodinger equation and, then, present the essential aspects of quantum dynamics by considering a series of particle scatterings and bound particle states. In the Discussion section we analyze the differences between our approach and the traditional one.

## II. PARTICLE WAVES AND WAVEFUNCTION

A particle in a single momentum-energy state does not oscillate. But an interacting particle has alternative intermediate states that interfere like waves and yield the probabilities of the various outcomes. We address this situation by associating each particle state (**p**,$E$) with *two* plane waves – one real and one imaginary – of the form

$$\psi_p(\mathbf{r},t) = a(\mathbf{p},E)e^{\frac{i}{\hbar}(p \cdot r - Et)} \tag{1}$$

The intensity $|\psi_p(\mathbf{r},t)|^2 = |a(\mathbf{p},E)|^2$ of a particle wave is constant in space and time, but two or more particle states interfere like waves.

In one dimension, we can visualize a particle wave by imagining at each point a clock with one hand rotating with period $T = h/E$, while the phase angle increases in the direction of the momentum attaining the same orientation $\lambda = h/p$ meters farther. In three dimensions, the plane wave fills the entire space and the phases of the "clock hands" increase in the direction of the momentum, completing a full cycle every $\lambda$ meters.

The particles we are interested in have positions and times as well as momenta and energies within certain ranges and are described by wavefunctions $\psi(\mathbf{r},t)$ that are superpositions of particle waves.

$$\psi(\mathbf{r},t) = \frac{1}{(2\pi\hbar)^{4/2}} \int a(\mathbf{p},E)e^{\frac{i}{\hbar}(p \cdot r - Et)} d^3\mathbf{p}dE$$

$$a(\mathbf{p},E) = \frac{1}{(2\pi\hbar)^{4/2}} \int \psi(\mathbf{r},t)e^{-\frac{i}{\hbar}(p \cdot r - Et)} d^3\mathbf{r}dt \tag{2}$$

The complete phase exponent in (2) is [**p**·**r**−$Et$+$\theta$(**p**,$E$)], however we can have $\theta$(**p**,$E$)=0 by choosing appropriately the origin of coordinates.

A detection of a particle will locate it at point (**r**,$t$) with probability $|\psi(\mathbf{r},t)|^2$, while a measurement of momentum-energy will find the particle in some state (**p**,$E$) with probability $|a(\mathbf{p},E)|^2$. From the wavefunction $\psi(\mathbf{r},t)$ and the amplitudefunction $a(\mathbf{p},E)$ we can calculate the total values of quantities that depend on space-time and on energy-momentum respectively, as in equations (4) and (16) below.

## III. DERIVATION OF THE SCHRODINGER EQUATION

A particle carrying energy $E$ and interacting with a time-independent potential $V(\mathbf{r})$ has specific total potential energy (*PE*) and, consequently, specific total kinetic energy (*KE*). All intermediate waves of this particle have energy $E$, and we can extract the time-energy dependence of the wavefunction as an unobservable phase factor. The space part $\psi(\mathbf{r})$ of the wavefunction is a superposition of wave states with amplitudes $a(\mathbf{p})$.

$$\psi(\mathbf{r}) = \frac{1}{(2\pi\hbar)^{3/2}} \int a(\mathbf{p})e^{\frac{i}{\hbar}p \cdot r} d^3\mathbf{p} \tag{3}$$

The description of the interaction by a potential introduces the energy equation that the total energy of the particle is equal to its total kinetic plus potential energies.

$$E = KE + PE \quad \rightarrow \quad E = \int |a(\mathbf{p})|^2 \frac{\mathbf{p}^2}{2m} d^3\mathbf{p} + \int |\psi(\mathbf{r})|^2 V(\mathbf{r}) d^3\mathbf{r} \tag{4}$$

Since $\psi(\mathbf{r})$ and $a(\mathbf{p})$ are interconnected through equation (3), the potential and kinetic energies are interrelated and the energy equation is fulfilled by a specific wavefunction only.

In particle scatterings, the particle energy ($E$) and the interaction potential $V(\mathbf{r})$ determine the wavefunction $\psi(\mathbf{r})$ − the unique superposition of wave-states for which the kinetic plus potential energies equal to $E$. For bound particles, the potential $V(\mathbf{r})$ plus boundary conditions determine the possible eigenfunctions $\psi_n(\mathbf{r})$ that fulfill the energy equation, and yield the energy eigenvalues $E_n$.

Using equation (3) we have

$$\int |a(\mathbf{p})|^2 \frac{\mathbf{p}^2}{2m} d^3\mathbf{p} = \int \psi^* \left[ -\frac{\hbar^2}{2m} \nabla^2 \psi \right] d^3\mathbf{r} \tag{5}$$

and substituting (5) into (4), we put the energy equation in the following form

$$\int \psi^* \left[ E\psi - \frac{\hbar^2}{2m} \nabla^2 \psi - V(\mathbf{r})\psi \right] d^3\mathbf{r} = 0 \tag{6}$$

The above equation states that the average value of the integrand is zero. But, as we will justify below, the integrand itself is zero at each point and, dividing it by $\psi^*$, we get

$$-\frac{\hbar^2}{2m} \nabla^2 \psi + V(\mathbf{r})\psi = E\psi \quad \text{at each } \mathbf{r} \tag{7}$$

This is the Schrodinger equation, derived as the unique condition the wavefunction must satisfy in order to fulfill the energy equation. Only a wavefunction that satisfies this condition produces the kinetic and potential energies that add up to the total energy.

The integrand $I(\mathbf{r})$ in (6) must vanish everywhere, because if it differs from zero at some point that would alter the value of the potential. For example, suppose that the integrand at point $\mathbf{r}$ is $I(\mathbf{r}) = N(\mathbf{r}) = \psi^* n(\mathbf{r}) \psi$. Then, we have

$$I(\mathbf{r}) = \psi^* \left[ E\psi + \frac{\hbar^2}{2m} \nabla^2 \psi - V(\mathbf{r})\psi \right] = \psi^* n(\mathbf{r})\psi$$

$$I'(\mathbf{r}) = \psi^* \left[ E\psi + \frac{\hbar^2}{2m} \nabla^2 \psi - \{V(\mathbf{r}) + n(\mathbf{r})\}\psi \right] = 0 \tag{8}$$

The value of the potential changes, arbitrarily, from $V(\mathbf{r})$ to $V(\mathbf{r})+n(\mathbf{r})$ and the wavefunction changes too. The amplitudes of the intermediate states of the new wavefunction depend on $V(\mathbf{r})$ and $n(\mathbf{r})$ – a local change of the potential generates a global change in the wavefunction. We get the wavefunction for the given potential for $n(\mathbf{r})=0$, therefore the Schrodinger equation [$I(\mathbf{r})=0$] is the unique physical condition that satisfies equation (6).

A classical potential is a strict but effective description of the interaction. Even though quantum energy fluctuations do happen, the potential is the average over these fluctuations. We may say that, in the non-relativistic theory, the particle "sees" only the classical potential and the superposition of its intermediate states "sculpts" the wavefunction that yields the proper potential and kinetic energies.

When the potential $V(\mathbf{r},t)$ depends on space *and* time, following similar steps, we derive from the energy equation the time-dependent Schrodinger equation

$$KE + PE = E \quad \rightarrow \quad -\frac{\hbar^2}{2m}\nabla^2\psi + V(\mathbf{r},t)\psi = i\hbar\frac{\partial \psi}{\partial t} \qquad (9)$$

where the wavefunction $\psi(\mathbf{r},t)$ is a superposition of particle wave-states having different momenta and energies.

## IV. QUANTUM DYNAMICS

Having derived the Schrodinger equation and, thus, removed its axiomatic status, the dynamics of quantum mechanics can be unified under one principle: The probabilities of physical outcomes are obtained from the superposition of the intermediate particle waves and interaction processes, considered as happening concurrently.

As we will see in the next two sections, all kinds of interactions determine what intermediate states may happen − i.e. interaction with boundaries or with classical potentials or quantum interactions processes. Different but overlapping dynamical descriptions yield identical results.

## V. PARTICLE SCATTERINGS

In particle scatterings, we determine experimentally the particle wavefunction by measuring the percentages of the various outcomes; we also derive the wavefunction, theoretically, by figuring out what intermediate states arise.

**Double Slit Experiment:** The interference pattern produced in a double slit experiment is the clearest demonstration that, even though the particle passes through one slit, the probabilities of the various outcomes are obtained from the superposition of all its intermediate waves. In particular, the wavefunction at each point on the film behind the slits arises from the interference of two intermediate particle waves emerging from the slits and meeting at that point.

**Light Reflection and Refraction:** In light reflection and refraction, a photon may travel between any two points along infinite alternative paths. However, by considering *strands* of neighboring paths, the travel times within each strand vary and the superposition of the photon waves produces a vanishing wavefunction; consequently, the photon cannot travel along these routes. However, the waves near the fastest path have equally fast travel times and interfere constructively, producing a strong wavefunction that the photon does follow.

**Tunneling:** When a particle wave hits a rectangular potential barrier, we know the forms of the reflected wave, the wave inside the barrier, and the transmitted wave. By requiring that these waves are simultaneously continuous at the edges of the barrier, we can determine the probabilities of the reflected and transmitted states.

**Electron-Nucleus Scattering:** The formal theory of particle scattering is based on the time-dependent Schrodinger equation but here we will follow a quantum interactions approach. In low energy electron-nucleus collisions, we can consider the nucleus (thousands of times more massive than the electron) as the fixed center of scattering. In this approximation, the energy of the scattered electron is equal to that of the incoming electron, and we obtain the first order scattering amplitude as follows.

The outgoing electron state in some angle $\theta$ is the sum of the contributions of all intermediate scattering events that produce electron waves of momentum $\mathbf{p}_\theta$. In such a scattering event at point $\mathbf{r}$ from the nucleus, the incoming electron state $\psi_0(\mathbf{r})$ interacts with the electric potential $V(r)$ ($=-kZe^2/r$) and scattered waves $\psi_{sc}(\mathbf{r})$ are radiated in all directions. The amplitudes of the scattered waves are proportional to the incoming wave and to the strength of the electric potential of the nucleus at $\mathbf{r}$

$$\psi_{sc}(\mathbf{r}) = V(r)\psi_0(\mathbf{r}) \tag{10}$$

The amplitude $a(\mathbf{p}_\theta)$ of an outgoing electron state is obtained by summing over all scattering events, considered as happening concurrently, and "projecting" $\psi_{sc}(\mathbf{r})$ onto $\psi_\theta$.

$$a(\mathbf{p}_\theta) \propto \int \psi_\theta^*(\mathbf{r})\psi_{sc}(\mathbf{r})d^3\mathbf{r} = \int e^{-\frac{i}{\hbar}p_\theta \cdot r} \times \frac{kZe^2}{r} \times e^{\frac{i}{\hbar}p_0 \cdot r} d^3\mathbf{r} \tag{11}$$

Carrying out the space integration, we obtain the well-known result

$$probability = |a(p_\theta)|^2 = \left[\frac{kZe^2}{4E\sin^2(\theta/2)}\right]^2 \tag{12}$$

In quantum interactions theory, each electron scattering event happens due to the exchange of one intermediate photon between electron and nucleus. An intermediate photon of momentum $q$ absorbed or emitted by the electron at point $\mathbf{r}$ contributes a factor $1/q^2$ to the scattering amplitude (a contribution that is inversely proportional to the deviation of the intermediate photon from its real state). Summing first over $q$, we get the integral expression in (11), an indication that the classical electric potentials arise from the exchanges of low energy intermediate photons.

## VI. BOUND PARTICLES

We determine experimentally the wavefuction of a bound particle by detecting the intermediate particle states. We derive the wavefunction theoretically either by figuring out what intermediate states arise or from the Schrodinger equation.

**A. Particle in a Box**

Consider an electron inside a one-dimensional, $L$ long box. We can obtain the lowest energy electron eigenfunction in two ways:

*Experimentally*: Measurements would show that the electron moves half the time to the right and half the time to the left with momentum $p_1 = h/(2L)$. These are the two intermediate electron states and the sum of their waves yields the electron eigenfunction.

*Theoretically*: Since the average electron momentum is zero, the electron eigenstate is the superposition of two intermediate states having opposite momenta. Furthermore, the eigenfunction must vanish at the walls of the box and the longest wavelength (lowest momentum) for which this can happen is $\lambda_1 = 2L$, corresponding to momenta $\pm h/(2L)$.

In general, every eigenfunction of a particle in the box has the form

$$\psi_n(x,t) = a(+p_n)e^{\frac{i}{\hbar}(p_n x - E_n t)} + a(-p_n)e^{\frac{i}{\hbar}(-p_n x - E_n t)} = \sqrt{\frac{2}{L}} \sin\frac{p_n}{\hbar}x \otimes e^{-\frac{i}{\hbar}E_n t} \quad (13)$$

$$\text{where} \quad p_n = np_1 = n\frac{h}{\lambda_1} = n\frac{h}{2L} \quad \text{and} \quad a(+p_n) = -a(-p_n) = \frac{-i}{\sqrt{2L}}$$

The signs and magnitudes of the amplitudes $a(+p_n)$ and $a(-p_n)$ are determined by the boundary conditions $\psi(0,t) = \psi(L,t) = 0$ and by the normalization requirement that the total probability is 1.

**B. Ground State of Hydrogen Atom**

We can detect the intermediate electron momentum states and measure their amplitudes $a_1(\mathbf{p})$ by hitting Hydrogen atoms with energetic photons. Such "photoelectric" experiments have been carried out with Hydrogen and Helium atoms.[3] These measurements yield that the probability of finding the electron in an intermediate momentum state $p$ is

$$\text{momentum probability} = |a_1(p)|^2 = \frac{1}{\pi^2}\left(\frac{2r_0}{\hbar}\right)^3\left[1 + \frac{r_0^2}{\hbar^2}p^2\right]^{-4} \quad (14)$$

where $r_0 = 0.053$ nm.

Inserting $a_1(p)$ into (3) we obtain the eigenfunction $\psi_1(r)$ and we get that the probability of finding the electron at distance $r$ from the nucleus is

$$\text{position probability} = |\psi_1(r)|^2 = \left[\frac{1}{\sqrt{\pi r_0^3}}e^{-\frac{r}{r_0}}\right]^2 \quad (15)$$

All intermediate electron states on the ground level have energy $E_1$, equal to the total kinetic plus potential energies which we calculate from $|a_1(p)|^2$ and $|\psi_1(r)|^2$.

$$E_1 = \int |a_1(p)|^2 \frac{\mathbf{p}^2}{2m}d^3\mathbf{p} + \int |\psi_1(r)|^2 \frac{-ke^2}{r}d^3\mathbf{r} = +13.6 - 27.2 = -13.6 \; eV \quad (16)$$

We also derive the above quantities theoretically as follows: An electron eigenstate is a superposition of intermediate waves that have energy $E_n$. Since there are no boundaries, intermediate states of all momenta (from $-\infty$ to $+\infty$) arise in every direction. An eigenfunction $\psi_n(\mathbf{r})$ must vanish at infinity and, in order to fulfill the energy equation, must satisfy at each point $\mathbf{r}$ the Schrodinger equation

$$-\frac{\hbar^2}{2m}\nabla^2 \psi_n - \frac{ke^2}{r}\psi_n = E_n \psi_n \tag{17}$$

For the ground electron state we have

$$\psi_1(\mathbf{r},t) = \psi_1(\mathbf{r}) e^{-\frac{i}{\hbar}E_1 t} \qquad \psi_1(\mathbf{r}) = \int_{-\infty}^{+\infty} a_1(\mathbf{p}) e^{\frac{i}{\hbar}\mathbf{p}\cdot\mathbf{r}} d^3\mathbf{p} \tag{18}$$

Solving equation (17) we obtain $\psi_1(\mathbf{r})$ and then we calculate $a_1(\mathbf{p})$ from equation (3). The calculated results agree with the experimental ones shown in (15) and (14), where now we get that $r_0 = \hbar^2/(mke^2) = 0.053$ nm and $E_1 = -ke^2/(2r_0) = -13.6$ eV.

## VII. ANGULAR MOMENTUM

Classical angular momentum terms like $xp_y$ are not meaningful since at each point $x$ the intermediate states of a particle have different $p_y$ momenta. But the following expression

$$x\hat{p}_y \psi = x\left(-i\hbar \frac{\partial}{\partial y}\right)\psi = \frac{1}{(2\pi\hbar)^{3/2}}\int x p_y \otimes a(\mathbf{p}) e^{\frac{i}{\hbar}\mathbf{p}\cdot\mathbf{r}} d^3\mathbf{p} \tag{19}$$

represents accurately the angular momentum term since $x$ is multiplied by $p_y$ times the probability amplitude for $p_y$. In the above expression, we have omitted the time-energy phase factors because do not contribute to the value of the angular momentum, but we have retained the space-momentum factors that are part of the amplitude. Thus, in quantum mechanics, the values of angular momenta are obtained from the angular momentum operators $\hat{L}_i = \varepsilon_{ijk} r_j \hat{p}_k$.

When the interaction potential is symmetric, the form of the particle wavefunction reflects this symmetry. For example, since the electric potential of the nucleus is spherically symmetric, the atomic electron eigenfunctions have the form

$$\psi_{nlm}(\mathbf{r}) = R_n(r)\Theta_l(\theta) e^{im\varphi} \tag{20}$$

where $r$, $\theta$, $\varphi$ are the spherical coordinates and $m$, $l$ are integers with $m = -l, -l+1, \ldots, 0, \ldots, l$.

In spherical coordinates, the angular momentum operators are

$$\hat{L}_z = -i\hbar \frac{\partial}{\partial \varphi}$$

$$\hat{L}^2 = -\hbar^2 \left[ \frac{1}{\sin\theta} \frac{\partial}{\partial \theta}\left(\sin\theta \frac{\partial}{\partial \theta}\right) + \frac{1}{\sin^2\theta} \frac{\partial^2}{\partial \varphi^2} \right]$$

$$\hat{L}_x = i\hbar \left[ \sin\varphi \frac{\partial}{\partial \theta} + \cot\theta \cos\varphi \frac{\partial}{\partial \varphi} \right]$$

$$\hat{L}_y = i\hbar \left[ -\cos\varphi \frac{\partial}{\partial \theta} + \cot\theta \sin\varphi \frac{\partial}{\partial \varphi} \right]$$

(21)

The total values of $L_z$ and $L^2$ for a particle described by $\psi_{nlm}(\mathbf{r})$ shown in (20) are

$$L_z = \int \psi_{nlm}^* \hat{L}_z \psi_{nlm} d^3\mathbf{r} = \hbar m$$

$$L^2 = \int \psi_{nlm}^* \hat{L}^2 \psi_{nlm} d^3\mathbf{r} = \hbar^2 l(l+1)$$

(22)

The total values of $L_x$ and $L_y$ are zero, their terms being proportional to $\cos\varphi$ and $\sin\varphi$. However, since the $x$, $y$, $z$ axes can be in any direction, the zero values of $L_x$ and $L_y$ indicate that once the $L_z$ has a specific eigenvalue, $L_x$ and $L_y$ cannot have specific, other than zero, total values. The above results provide the physical basis for the theory of angular momentum in quantum mechanics.

## VIII. TIME VARIATION

The interdependence between time ($t$) and energy ($E$) is analogous to that between position ($\mathbf{r}$) and momentum ($\mathbf{p}$). Thus, while intermediate states having different momenta yield wavefunctions that vary in space, intermediate states with different energies produce wavefunctions that vary in time. For example, wavefunctions that are superpositions of two or more energy eigenfunctions of a particle in a box, a harmonic oscillator, an atomic electron, etc. vary in space *and* time.

When the interaction is described by a potential $V(\mathbf{r},t)$, which depends explicitly on space and time, intermediate states having different energies arise and the wavefunction satisfies the time-dependent Schrodinger equation. Formally, we can write the energy equation and the corresponding wavefunction equation as follows

$$E = H \quad \Leftrightarrow \quad i\hbar \frac{\partial \psi}{\partial t} = \hat{H}\psi$$

(23)

where $H$ is the Hamiltonian of the system and $\hat{H}$ is the Hamiltonian operator. When the Hamiltonian has the form $H_0+H_I$, where we can solve the problem for $H_0$ while $H_I$ is a smaller interaction term, we can use the above equation to obtain approximate series solutions.

## IX. DISCUSSION

The fundamental principle of quantum mechanics is that the probabilities of physical outcomes are obtained from the intermediate particle waves and interaction processes, considered as happening concurrently. Quantum dynamics includes any kind of particle interaction, i.e. boundaries, potentials, quantum interactions. We have seen how boundaries determine the intermediate states in the double slit experiment, light reflection, light refraction, tunneling, and particle in a box. In quantum interactions theory, i.e. Compton scattering as described by quantum electrodynamics, the wavefunction is obtained by adding together all intermediate interaction processes.

When the interaction is described by a classical potential, the wavefunction must fulfill the energy equation and, as we have shown, it must satisfy the Schrodinger equation at all points. Likewise, we can derive that the Dirac equation − for an electron interacting with an electromagnetic potential − is the unique condition the wavefunction must satisfy at each point in order to fulfill the relativistic energy equation.[4] The Pauli equation, which includes the interaction of the electron spin with the magnetic field, is obtained as a low energy approximation of the Dirac equation or by "linearizing" the Schrodinger equation.[5]

Our definition of the wavefunction shows the physical content of the angular momentum operators and justifies their form. Wavefunctions of two or more particles, spin states, etc. are formulated in the usual way.

In the traditional approach, the foundation of non-relativistic quantum theory is the postulated Schrodinger equation, which we solve to get the wavefunctions of a particle in a box, an atomic electron or an electron colliding with a nucleus. However, we miss the tangible physical descriptions and explanations present in our treatment of these problems.

Another notable difference is the *local* versus *global* nature of quantum theory. The traditional attitude is to proceed in analogy to classical physics and find an "equation of motion", which is the axiomatically introduced Schrodinger equation; thus, quantum mechanics appears to be a local space-time theory. In our approach, non-relativistic quantum dynamics has a space-time global nature because particle states have constant amplitudes. In particular, since the wavefunction of two particles consists of waves with constant amplitudes, a measurement would detect one of the possible intermediate two particle states, and the particles would appear to be "entangled".

Here, a brief discussion of the history and status of the Schrodinger equation is in order. Erwin Schrodinger obtained this equation in January 1926, before the realization of the probabilistic nature of particle waves. Schrodinger's original derivation has been considered an "invention" because no solid, physically justified derivation was ever found. As a consequence, current textbooks regard the Schrodinger equation an axiom introducing it with statements like "We cannot derive the Schrodinger equation from more basic principles, it is the basic principle".[6]

Now, multiplying the Schrodinger equation by $\psi^*$ and integrating over space and time we obtain the energy equation; but it is not obvious that the reverse would hold. Nevertheless, logic and physical perception suggest that there must be a unique interconnection between the two equations − and this is what we have shown. The Schrodinger equation has the appearance of an "equation of motion", but it is the local condition the wavefunction must satisfy in order to fulfill the overall energy equation.

The unified description of quantum dynamics we have presented yields a tangible, experimentally justified and theoretically consistent quantum theory.